\documentstyle{article}
\begin{document}
\newcommand{\beq}{\begin{equation}}
\newcommand{\eeq}{\end{equation}}
\newcommand{\beqn}{\begin{eqnarray}}
\newcommand{\eeqn}{\end{eqnarray}}
\newcommand{\dpf}{\displaystyle\frac}
\newcommand{\no}{\nonumber}
\newcommand{\ep}{\epsilon}
\begin{center}
{\Large Entropy bound for a rotating system from Anti de Sitter black holes}
\end{center}
\vspace{1ex}
\centerline{\large Bin
Wang$^{a,b,}$\footnote[1]{e-mail:binwang@fma.if.usp.br},
\ Elcio Abdalla$^{a,}$\footnote[2]{e-mail:eabdalla@fma.if.usp.br}}
\begin{center}
{$^{a}$ Instituto De Fisica, Universidade De Sao Paulo, C.P.66.318, CEP
05315-970, Sao Paulo, Brazil \\
$^{b}$ Department of Physics, Shanghai Teachers' University, P. R. China}
\end{center}
\vspace{6ex}
\begin{abstract}
General geodesic equations of the motion of spinning systems around the (3+1)-dimensional
and (2+1)-dimensional rotating anti-de Sitter black holes have been obtained. 
Based upon these equations, we derived the entropy bound for a rotating
system from Kerr-Anti de Sitter black holes and BTZ black holes, respectively. Our result
coincides with that of Hod's derived from Kerr black hole, which shows that the entropy
bound of the rotating system is neither dependent on the black
hole parameters, nor on 
spacetime dimensions. It is a universal entropy bound. 
\end{abstract} \vspace{6ex} \hspace*{0mm} PACS number(s): 04.70.Dy,
05.70.Ce, 95.30Tg, 97.60.Lf.
\vfill
\newpage
\section{Introduction}

In order to rescue the generalized second law (GSL) of thermodynamics, Bekenstein
conjectured, some time ago, that there exists an upper bound on the entropy of any
neutral
object of energy $E$ and maximal radius $R$ in the form $S\leq 2\pi ER/\hbar$ [1]. This
derivation was criticized by Unruh, Wald and Pelath [2-4] for neglecting
the effects of
buoyancy in acceleration radiation. They concluded that no additional assumption for
upper entropy bound is necessary to maintain the GSL. However their criticism was refuted
by Bekenstein who showed that the buoyancy can really be negligible and does not spoil
the entropy bound derivation [5,6]. Bekenstein's entropy bound has received independent
support [7-10]. 

Recently, extending the derivation of an upper bound on the entropy to any charged
object, 
Bekenstein and Mayo [11], Hod[12] and Linet[13] have shown that Bekenstein's original
entropy bound can be improved. A tighter entropy bound
for the nonrotating object of mass $\mu$, radius $R$ and charge $e$
is required. It has been shown that such a bound is $S\leq
(2\pi R/\hbar)(E^2-e^2/2R)$. This
result agrees to an earlier finding by Zaslavskii [14] in another context.
However, the fact that this entropy bound for a charged system is necessary to uphold the
GSL has been challenged as well [15]. 

A tighter bound on entropy for objects with angular momentum has also been derived
recently [16]. Refering to Hojman and Hojman's [17] integrals of motion for a neutral
object with spin $s$ moving on a Kerr black hole background, Hod obtains the entropy
bound $S\leq \dpf{2\pi ER}{\hbar}(1-s^2/E^2 R^2)^{1/2}$. He claimed that this bound is
universal and independent of the black hole parameters which were used to derive it. In
order to examine this argument, in this paper we will study the entropy bound for a
spinning object falling into Anti de Sitter (AdS) black holes including
(3+1)-dimensional Kerr-AdS black holes and (2+1)-dimensional BTZ black holes. We will
derive geodesic equations of the motion of the spinning objects around Kerr-AdS black
holes and BTZ black holes, respectively. Based upon these geodesic equations, we will
show that the entropy bound for a rotating system depends neither on the black hole
parameters, nor on spacetimes dimensions. It is a universal result.

\section{Entropy bound from the Kerr-AdS black holes}

Recently the study of the Kerr-AdS black hole model has been undertaken and shown to be
important in many aspects[18-21]. The
metric is
\beqn         
{\rm d}s^2 & = &-\dpf{\Delta_r}{\rho^2}[{\rm d}t-\dpf{a}{\Sigma}\sin^2\theta{\rm
d}\phi]^2+\dpf{\rho^2}{\Delta_r}{\rm d}r^2+\dpf{\rho^2}{\Delta_{\theta}}{\rm d}\theta^2
\\ \no
&   &+\dpf{\sin^2\theta\Delta_{\theta}}{\rho^2}[a{\rm d}t-\dpf{(r^2+a^2)}{\Sigma}{\rm
d}\phi]^2
\eeqn
where
\beqn     
\rho^2 & = & r^2+a^2\cos^2\theta    \\ \no
\Delta_r & = & (r^2+a^2)(1+l^2 r^2)-2Mr  \\ \no
\Delta_{\theta} & = & 1-l^2a^2\cos^2\theta \\ \no
\Sigma & = & 1-l^2a^2
\eeqn
The parameter $M$ is the mass, $a$  the angular momentum per
unit mass, and
$l^2=-\Lambda/3$, where $\Lambda$ is the (negative) cosmological constant. The black hole
solution is valid for $a^2<l^{-2}$ [18-21]. For $l=0$, eq.(1) goes back to the metric for
Kerr
black hole. There are four roots of the polynomial $\Delta_r$, the largest root $r_+$
corresponds to the event horizon, the other positive root $r_-$ is the Cauchy horizon and
another two roots $r_1, r_2$ are negative and satisfy $r_1+r_2=-(r_+ + r_-), r_1
r_2=\dpf{a^2}{l^2r_+ r_-}$. The mass $M$, and the angular momentum per unit mass $a$, can
both be
expressed in terms of  $r_+, r_-$ and $l$ as
\beqn    
M & = & \dpf{(1+l^2r_-^2)(r_+ +r_-)(1+l^2r_+^2)}{2(1-l^2r_+ r_-)}, \\
a & = & \sqrt{\dpf{r_+ r_-(1+l^2r_-^2+l^2r_+ r_- +l^2r_+^2)}{1-l^2r_+ r_-}}.
\eeqn
The above equations require $l^2<1/r_+ r_-$ to ensure real and positive values of $a$ and
$M$, respectively.
For the extreme black hole case $r_+$ and $r_-$ degenerate and $M=M_e$, where $M_e$ is
the critical mass parameter given in [18]. 

We consider a spinning object of rest mass $m$, intrinsic spin $s$ and proper cylindrical
radius $R$, which is descending into the Kerr-AdS black hole. Following [17], the
constants of motion associated with the $t$ and $\phi$ variables are
\beqn       
E & = & \pi_t-g_{t\phi,r}\pi_t\dpf{s\Sigma}{2rm}+g_{tt,r}\pi_{\phi}\dpf{s\Sigma}{2rm} \\
J & = & -\pi_{\phi}-g_{\phi
t,r}\pi_{\phi}\dpf{s\Sigma}{2rm}+g_{\phi\phi,r}\pi_t\dpf{s\Sigma}{2rm}
\eeqn
where
\beqn     
\pi_t=g_{tt}\dot{t}+g_{t\phi}\dot{\phi} \\
\pi_{\phi}=g_{t\phi}\dot{t}+g_{\phi\phi}\dot{\phi}
\eeqn
For simplicity we just consider the equatorial motions of the object. The quadratic
equation for the conserved energy $E$ of the body is
\beq    
\tilde{\alpha}E^2-2\tilde{\beta}E+\tilde{\gamma}=0
\eeq
where the very long expressions for $\tilde{\alpha}, \tilde{\beta}, \tilde{\gamma}$ are
given in the
appendix. It is worth noting that taking $l\rightarrow 0$, they reproduce the 
expressions
given in [17]. 

In the spirit of the analysis of Bekenstein and of Hod, we neglect the buoyancy
contribution, for
simplicity. Suppose the gradual approach to the black hole must stop when the proper
distance from the body's center of mass to the black hole horizon equals $R$, the body's
radius
\beq       
\int_{r_+}^{r_+ +\delta(R)}(g_{rr})^{1/2} dr =R,
\eeq
with $g_{rr}=\dpf{r^2}{\Delta_r}$ (in equatorial plane) and
$\Delta_r=l^2(r-r_+)(r-r_-)(r-r_1)(r-r_2)$. One can get from integrating Eq.(10)
\beq       
\delta(R)=\dpf{l^2(r_+ -r_-)(r_+ -r_1)(r_+ -r_2)R^2}{4r_+^2}.
\eeq
Considering the relation between $r_1, r_2$ and $r_+, r_-$ together with
eqs.(3,4), it is
not difficult to find that when $l\rightarrow 0$, eq.(11) reproduces 
eq.(6) of [16].

Using the test particle approximation $s/(mr_+)\ll 1, R\ll r_+$ together with the
condition $l^2<1/(r_+r_-)$, we can solve eq.(9) for $E$ to first order in the small
quantities at the point of capture $r=r_+ +\delta(R)$,
\beq          
E=u+vs+R\sqrt{w}
\eeq
where
\beqn    
u & = & \dpf{aJ(a^2 l^2-1)(2M-a^2 l^2 r_+ -l^2r_+^3)}{-2a^2 M-a^2 r_+ +a^4 l^2 r_+
-r_+^3+a^2 l^2 r_+^3} \\
v & = &  \dpf{J(a^2 l^2 -1)}{mr_+(2a^2 M+a^2 r_+-a^4l^2 r_++r_+^3-a^2l^2r_+^3)^2} \\ \no
  &   & \times(2a^4 M-6a^2M^2r_++3a^2Mr_+^2 +5 a^4l^2Mr_+^2 -a^2 r_+^3+2a^4l^2r_+^3 \\
\no
  &   & -a^6l^4r_+^3+3Mr_+^4+3a^2l^2Mr_+^4-r_+^5+2a^2l^2r_+^5-a^4l^4r_+^5)\\ \no
w & = & \dpf{(r_+ -r_-)^2(-1-l^2r_-^2-2l^2r_- r_+-3l^2r_+^2+l^4r_-^2r_+^2+2l^4r_-
r_+^3)^2}{4(1+l^2r_-^2)^4 r_+^2(r_- +r_+)^4} \\ 
  &   & \times(J^2+m^2r_-^2+2l^2m^2r_-^4+l^4m^2r_-^6-4J^2l^2r_- r_+ +2m^2r_- r_+
-2J^2l^4r_-^3r_+ \\ \no
  &   & +4l^2m^2r_-^3r_+ + 2l^4m^2r_-^5 r_+
+m^2r_+^2+2J^2l^4r_-^2r_+^2+2l^2m^2r_-^2r_+^2+4J^2l^6r_-^4r_+^2 \\ \no
  &   & +l^4m^2r_-^4r_+^2+J^2l^8r_-^6r_+^2-2J^2l^4r_-
r_+^3+4J^2l^6r_-^3r_+^3+2J^2l^8r_-^5r_+^3+4J^2l^6r_-^2r_+^4\\ \no
  &   & +3J^2l^8r_-^4r_+^4+2J^2l^8r_-^3r_+^5+J^2l^8r_-^2r_+^6).
\eeqn
Eq.(12) reduces to (7) of [16] after taking the limit $l\rightarrow 0$ and substitution
of Eqs.(3,4).

After the assimilation of the spinning body, the change of the black hole mass and
angular momentum are $dM=E$ and $dL=J$, respectively. Taking cognizance of eq.(12) and
using
the first-law of black hole thermodynamics,
\beq     
dM=\dpf{\kappa}{8\pi}dA+\Omega dL
\eeq
where $\kappa=\dpf{r_+(1+a^2l^2+3l^2r_+^2-a^2/r_+^2)}{2(r_+^2+a^2)}$ and
$\Omega=\dpf{a(1-l^2a^2)}{r_+^2+a^2}$ are the surface gravity and rotational angular
frequency of the black hole respectively, we find  
\beq   
dA=\dpf{8\pi}{\kappa}(u+vs+R\sqrt{w}-\Omega J)
\eeq
Substituting eqs.(3,4), it is easy to see that $u-\Omega J=0$. Carefully choosing the
total angular momentum of the body at the critical value
\beq          
J=J^*=\sqrt{\dpf{m^2(1+l^2r_-^2)^2(r_++r_-)^2s^2}{(-1+2l^2r_+r_-+l^4r_-^3r_++l^4r_-^2r_+^2+l^4r_-r_+^3)^2(m^2R^2-s^2)}},
\no
\eeq
the minimum value of the
increase in the
black hole surface area caused by an assimilation of a spinning body with given
parameters $m,s,R$ is
\beq         
dA_{min}=8\pi\sqrt{m^2R^2-s^2}
\eeq
The minimum exists only for $s\leq mR$.

By virtue of the GSL, we derived an upper bound to the entropy $S$ of an arbitrary
system of proper energy $E$, intrinsic angular momentum $s$ and proper radius $R$ falling
into the Kerr-AdS black hole
\beq   
S\leq 2\pi\sqrt{(RE)^2-s^2}
\eeq

It is evident that the entropy $S$ of the rotating system should be bounded and this
bound is more stringent than the original Bekenstein bound [1]. It is worth noting that
although we used a different black hole model to derive the entropy bound, the final
result is the same as in [16] and independent of the black hole parameters which are used
to derive it.

\section{Entropy bound from the BTZ black hole}

The entropy bound for the rotating system derived from the (3+1)-dimensional Kerr-AdS
black hole is the same as that from Kerr black hole [16], which shows that this bound
does not depend on the black hole parameters. Now it is of interest to ask
the question
whether this bound only exists for (3+1)-dimensional cases and whether it will be changed
if it is derived from a different dimensional black hole. In this section we will
concentrate our attention on the (2+1)-dimensional BTZ black holes [23,24]. The metric of
this black hole reads
\beq 
{\rm d}s^2=-N^2{\rm d}t^2+N^{-2}{\rm d}r^2+r^2(N^{\phi}{\rm d}t+{\rm d}\phi)^2 \\
\eeq 
where the squared lapse $N^2(r)$ and the angular shift $N^{\phi}(r)$ are
\beqn 
N^2        & = & -M+\dpf{r^2}{l^2}+\dpf{J ^2}{4r^2} \\ \no
N^{\phi}   & = & -\dpf{J}{2r^2}                           \no
\eeqn
where $-\infty<t<+\infty, 0<r<\infty$, and $0\leq\phi\leq 2\pi$.  $M, J$ appearing
in (22) are two constants of integration which can be interpreted as the
black hole
mass and angular momentum. The lapse function $N^2(r)$ vanishes when
\beq                    
r_{\pm}=l\{\dpf{M}{2}(1\pm[1-(\dpf{J}{Ml})^2]^{1/2})\}^{1/2}
\eeq
Here $r_+$ is the event horizon and $r_-$ is the Cauchy horizon for $M>0$ and $\vert
J\vert<Ml$. In the extreme case $\vert J\vert= Ml$, $r_+$ and $r_-$ degenerate.

We proceed to devise a gedanken experiment in which a spinning object of mass $m$,
intrinsic spin $s$ and proper cylindrical radius $R$ is decending into the BTZ black
hole. Using constants of motion associated with the $t$ and $\phi$ definition
\beqn 
E & = & \pi_t-g_{t\phi,r}\pi_t\dpf{s}{2mr}+g_{tt,r}\pi_{\phi}\dpf{s}{2mr} \\ 
L & = & -\pi_{\phi}-g_{\phi t,r}\pi_{\phi}\dpf{s}{2mr}+g_{\phi\phi,r}\pi_t\dpf{s}{2mr} 
\eeqn
We obtained the quadratic equation for the conserved energy $E$ of the spinning body in
the form of (9), here 
\beqn  
\alpha & = & r^2-Js/m+(M-r^2/l^2)s^2/m^2 \\
\beta & = & JL/2 -LMs/m+JLs^2/(2l^2m^2) \\
\gamma & = & L^2M-J^2m^2/4-L^2r^2/l^2+Mm^2r^2-m^2r^4/l^2-JL^2s/(l^2m) \\ \no
       &   &
+[J^2/2l^2-2Mr^2/l^2+L^2r^2/(l^4m^2)+2r^4/l^4]s^2-(J^2/4-Mr^2+r^2/l^2)s^4/(l^4m^2)
\eeqn
Taking $s\rightarrow 0$ and adopting re-scalings given in [25], the quadratic equation
reproduces the special case given in [25].

Neglecting the buoyancy contribution, we suppose the gradual approach of the spinning
object to the black hole must stop when the proper distance from the body's center of
mass to the black hole horizon equals $R$. Considering
$N^2=\dpf{(r^2-r_+^2)(r^2-r_-^2)}{l^2r^2}$, from Eq.(10), we have
\beq   
\delta(R)=\dpf{(r_+^2-r_-^2)R^2}{2r_+ l^2}.
\eeq
Using conditions of approximation for test particle $s/(mr_+)\ll 1, R\ll r_+$, we can get
the energy expression in the same form as (12), where $u,v,w$ here are
\beqn    
u & = & \dpf{JL}{2r_+^2} \\
v & = & \dpf{J^2L}{2mr_+^4}-\dpf{LM}{mr_+^2} \\
w & = & \dpf{(r_+-r_-)^2(r_++r_-)^2(L^2+m^2r_+^2)}{l^4r_+^4}.
\eeqn

After the infall of the spinning object, the change of the black hole mass and angular
momentum are $dM=E$ and $dL=L$, respectively. Employing the first-law of black hole
thermodynamics Eq. (16), where the surface gravity and angular velocity
here are
$\kappa=\dpf{\sqrt{M^2-J^2/l^2}}{r_+}=\dpf{r_+^2-r_-^2}{l^2r_+}, \Omega=\dpf{J}{2r_+^2}$.
When the total angular momentum attains the critical value
\beq      
L^*=\dpf{mr_+ s}{\sqrt{m^2R^2-s^2}},
\eeq
there is a minimum increase in the black hole surface area caused by an
assimilation of the spinning body in the form
\beq  
dA_{min}=8\pi\sqrt{m^2R^2-s^2}
\eeq
This minimum exists only for $s\leq mR$.

Arguing from the GSL, we derive an upper bound to the entropy $S$ of an arbitrary system
of proper energy $E$, intrinsic angular momentum $s$ and proper radius $R$ from the BTZ
black hole
\beq       
S\leq 2\pi\sqrt{(RE)^2-s^2}
\eeq

This result is the same as obtained from the (3+1)-dimensional cases.

\section{Conclusions and discussions}

Using the method proposed by Hojman and Hojman [17], we have derived the geodesic
equations for the spinning object moving around the (3+1)-dimensional Kerr-AdS black
holes and (2+1)-dimensional BTZ black holes, respectively. These geodesic equations are
general compared to those obtained in the models without cosmological constant in [17]
for Kerr black hole case and without considering the spinning of the object for BTZ black
hole [25]. Based upon these geodesic equations, we derived the entropy bound for the
rotating system to maintain the GSL. These results coincide with that obtained from Kerr
black hole [16], which supports Hod's argument that the entropy bound of the rotating
sysytem is independent of the black hole parameters. Besides it is worth noting that our
result from the three-dimensional BTZ black holes indicates that this entropy bound
is also independent of the dimensions of spacetimes. Therefore the entropy bound for the
rotating system is universal.

However at the first sight, the universality of the entropy bound for the rotating system
cannot be extended to a charged system. Because at least the electric potential will
change the form when we study it for the (2+1)-dimensional case. The exact dependence of
the entropy bound on the system's parameters for the charged system still needs further
exploration.

ACKNOWLEDGEMENT: This work was partically supported
by Fundac\~ao de Amparo \`{a} Pesquisa do Estado de
S\~{a}o Paulo (FAPESP) and Conselho Nacional de Desenvolvimento 
Cient\'{\i}fico e Tecnol\'{o}gico (CNPQ).  B. Wang would also
like to acknowledge the support given by Shanghai Science and Technology
Commission.
  
\section{Appendix}

Here we give the expressions of $\tilde{\alpha}, \tilde{\beta}, \tilde{\gamma}$ in eq.(9)
in the main text. 
\beqn            %
\tilde{\alpha} & = & \alpha+k_1l^2+k_2l^4 \\ \no
\tilde{\beta} & = & \beta+p_1l^2+p_2l^4+p_3l^6 \\ \no
\tilde{\gamma} & = & \gamma+q_1l^2+q_2l^4+q_3l^6+q_4l^8 \no
\eeqn
where 
\beqn          %
k_1& =& -a^4-a^2r^2+(a^2+2a^2M/r-r^2)s^2/m^2 \\ \no
k_2 & = & a^2r^2s^2/m^2 \\ \no
p_1& =& -a^3J-2a^3JM/r-aJr^2+(-2a^2J+2a^4JM/r^3+3a^2JM/r)s/m \\ 
   &  & +(aJ-a^3JM^2/r^4-a^3JM/r^3+2aJM/r)s^2/m^2 \\ \no
p_2& =& a^5J+a^3Jr^2+a^4Js/m+(-a^3J-2a^3JM/r+ aJr^2)s^2/m^2 \\ \no
p_3 & = & -a^3Jr^2s^2/m^2 \\ \no
q_1& =& a^2J^2-4a^2J^2M/r-J^2r^2-a^2m^2r^2-m^2r^4+(-2aJ^2+4a^3J^2M/r^3)s/m \\ 
   &  &+[2a^2-2a^2J^2M^2/(m^2r^4)+2a^2M/r+2J^2M/(m^2r)-2Mr+2r^2]s^2 \\ \no
   &  &+[-a^2M^2/r^4-2a^2M/r^3+3M^2/r^2-2M/r]s^4/m^2 \\ \no
q_2& =& a^4J^2+2a^4J^2M/r+2a^2J^2r^2+(4a^3J^2-2a^5J^2M/r^3)s/m \\ \no
   &  & +[a^4J^2M^2/(m^2r^4)-4a^2J^2M/(m^2r)+2a^2r^2+J^2r^2/m^2+2r^4]s^2 \\ \no
   &  & +(-a^2-2a^2M/r-r^2)s^4/m^2    \\ \no
q_3& =& -a^6J^2-a^4J^2r^2-2a^5J^2s/m \\ \no
   &  & +(2a^4J^2M/r-2a^2J^2r^2)s^2/m^2+(-a^2r^2-r^4)s^4/m^2 \\ \no
q_4& =& a^4J^2r^2s^2/m^2
\eeqn
$\alpha, \beta$ have the same expressions given in [17]. There is a sign mistake for the
expression $\gamma$ in [17], where the last term should be negative. The correct formula
is
\beq 
\gamma=-e^2\phi^2k_1+2e\phi(j-eh)k_3-(j-eh)^2k_2-\delta^2\Delta M^2 \no
\eeq
so that the requirement 
\beq
\beta^2-\alpha \gamma=\Delta\delta^2[(j-eh)^2-k_1M^2] \no
\eeq
can be satisfied.

\end{document}